\documentclass[conference,9pt]{IEEEtran}

\usepackage{amsmath,amssymb,amsfonts}
\usepackage{algorithmic}
\usepackage{graphicx}
\usepackage{textcomp}
\usepackage{booktabs} 
\usepackage[utf8]{inputenc}
\usepackage{listings}

\usepackage[sorting=none]{biblatex}
\AtBeginBibliography{\footnotesize}

\usepackage[per-mode=symbol]{siunitx}
\usepackage{stfloats}
\usepackage[caption=false,font=normalsize,labelfont=sf,textfont=sf]{subfig}
\def\BibTeX{{\rm B\kern-.05em{\sc i\kern-.025em b}\kern-.08em
    T\kern-.1667em\lower.7ex\hbox{E}\kern-.125emX}}
\usepackage{titlesec}
\usepackage{subcaption}
\usepackage{pifont}
\usepackage[x11names,dvipsnames]{xcolor}
\DeclareSIUnit{\cycle}{cycle}

\newcommand{\cmark}{\ding{51}} 
\newcommand{\xmark}{\ding{55}} 

\definecolor{mygreen}{HTML}{005f73}
\definecolor{myred}{HTML}{6f1a07}

\definecolor{keywordcolor}{RGB}{0, 116, 128}   
\definecolor{stringcolor}{RGB}{151, 102, 91}   
\definecolor{commentcolor}{gray}{0.5}
\definecolor{backgroundcolor}{RGB}{255, 255, 255}

\begin{document}

\title{Invited Paper: \mbox{FEMU}: An Open-Source and\\ Configurable Emulation Framework for Prototyping \\TinyAI Heterogeneous Systems}


\author{
\IEEEauthorblockN{Simone Machetti\IEEEauthorrefmark{1},
Deniz Kasap\IEEEauthorrefmark{1},
Juan Sapriza\IEEEauthorrefmark{1},
Rubén Rodríguez Álvarez\IEEEauthorrefmark{1},
Hossein Taji\IEEEauthorrefmark{1},\\
 Jos\'e Miranda\IEEEauthorrefmark{6},
Miguel Peón-Quirós\IEEEauthorrefmark{4},
David Atienza\IEEEauthorrefmark{1}} \\
\IEEEauthorblockA{\IEEEauthorrefmark{1}Embedded Systems Laboratory (ESL), EPFL, Lausanne, Switzerland}
\IEEEauthorblockA{\IEEEauthorrefmark{6}Center of Industrial Electronics (CEI), UPM, Madrid, Spain}
\IEEEauthorblockA{\IEEEauthorrefmark{4}EcoCloud, EPFL, Lausanne, Switzerland} \\
\IEEEauthorblockA{simone.machetti@epfl.ch, deniz.kasap@epfl.ch, juan.sapriza@epfl.ch, ruben.rodriguezalvarez@epfl.ch, hossein.taji@epfl.ch,\\ jose.miranda@upm.es, miguel.peon@epfl.ch, david.atienza@epfl.ch}
}

\maketitle

\begin{abstract}

    In this paper, we present the new FPGA EMUlation (\mbox{FEMU}), an open-source and configurable emulation framework for prototyping and evaluating TinyAI heterogeneous systems (HS). \mbox{FEMU} leverages the capability of system-on-chip (SoC)-based FPGAs to combine the under-development HS implemented in a reconfigurable hardware region (RH) for quick prototyping with a software environment running under a standard operating system in a control software region (CS) for supervision and communication. To evaluate our approach, we built the \mbox{X-HEEP} FPGA EMUlation (\mbox{X-HEEP-FEMU}) platform by instantiating the proposed framework with real-world hardware and software components. \mbox{X-HEEP-FEMU} is deployed on the Xilinx Zynq-7020 SoC and integrates the eXtendible Heterogeneous Energy Efficient Platform (\mbox{X-HEEP}) host in the RH, a Linux-based Python environment on the ARM Cortex-A9 CS, and energy models derived from a TSMC \SI{65}{\nano\meter} CMOS silicon implementation of \mbox{X-HEEP}, called \mbox{HEEPocrates}.\\

\end{abstract}


\section{INTRODUCTION}

    The increasing demand for real-time machine learning applications has driven the rapid growth of edge computing. By processing data directly on local devices instead of relying on remote cloud infrastructure, edge computing minimizes latency, strengthens data privacy, and enhances energy efficiency. These are key advantages for many edge-centric use cases. 

    However, these workloads impose strict requirements on processing power, responsiveness, and energy consumption, highlighting the need for customized hardware solutions~\cite{co-design_vision}. An effective strategy to meet these requirements is the use of heterogeneous systems (HS), which combine a general-purpose central processing unit (CPU) as host with specialized hardware accelerators to optimize both performance and energy efficiency~\cite{heepocrates, asip, e-gpu, taji2025medea}. The design of such systems is particularly critical for ultra-low-power edge applications (TinyAI), where balancing performance, power, and energy efficiency presents significant challenges.

    These applications run on battery-powered devices operating under tight power constraints, typically in the range of tens of milliwatts, necessitating highly efficient architectures. Their small form factor also imposes strict area restrictions, with a complete system-on-chip (SoC) often occupying only a few square millimeters. The challenges further increase on the software side, where applications generally involve two distinct, possibly overlapping, operational phases~\cite{bio_apps, ace}:~(1)~acquisition, where sensor data are collected through analog-to-digital converters (ADCs) and stored in memory~\cite{taji2024energy}, and~(2)~processing, where these data are analyzed.

    To efficiently prototype and evaluate such systems, tight hardware/software co-design is essential. This requires flexible and accurate development platforms capable of supporting real-time interaction, as well as performance and energy measurements. Existing register transfer level (RTL)-based platforms~\cite{pulp, opentitan} offer cycle-accurate modeling, but are slow, inflexible, and require complete hardware implementations, making them unsuitable for early-stage design exploration. On the other hand, software-based simulators~\cite{gem5} are easier to use and faster, but typically lack sufficient accuracy or realism to guide design decisions for TinyAI systems reliably.


    To overcome these challenges, we present FPGA EMUlation (\mbox{FEMU}), an open-source and configurable emulation framework for prototyping and evaluating TinyAI HS. This approach leverages the capability of SoC-based FPGAs to combine the under-development HS implemented in a reconfigurable hardware region (RH) for quick prototyping, with a software environment running under a standard operating system (OS) on a control software region (CS) for supervision and communication. HS peripherals are virtualized on the CS, enabling flexible communication and processing of comprehensive signal databases to evaluate the correctness of the system. Custom accelerators can be incorporated as virtualized software models in the CS for early-stage testing or as RTL modules in the RH for later-stage development. The framework supports system-level performance and energy estimation for combinations of virtualized and RTL accelerators using models derived from real-world silicon measurements. Developers interact with \mbox{FEMU} via a software interface, which simplifies communication and accelerates design space exploration.

    To demonstrate our approach, we implement the \mbox{X-HEEP} FPGA EMUlation (\mbox{X-HEEP-FEMU}) platform by instantiating the proposed framework with real hardware and software components. \mbox{X-HEEP-FEMU} is deployed on the Xilinx Zynq-7020 SoC and integrates the eXtendible Heterogeneous Energy Efficient Platform (\mbox{X-HEEP})~\cite{x-heep} host in the RH, a Linux-based Python environment on the ARM Cortex-A9 CS, and energy models derived from a TSMC \SI{65}{\nano\meter} CMOS silicon implementation of \mbox{X-HEEP}, referred to as \mbox{HEEPocrates}~\cite{heepocrates}.

    The key contributions of this work are:
    
    \begin{itemize}
        \item \mbox{FEMU}: a flexible and reusable emulation framework for prototyping and evaluating TinyAI HS.
        \item A hybrid SW/HW strategy that enables debugging, ADC, and flash virtualization, as well as accelerator integration during both early-stage development using software models and later-stage deployment with RTL implementations.
        \item A system-level approach for performance and energy estimation, grounded in real silicon measurements and enabling practical evaluation of HS.
        \item \mbox{X-HEEP-FEMU}: a complete emulation platform realized by instantiating the proposed framework with real-world hardware and software components, released as open source.\footnote{The complete code and documentation of the \mbox{X-HEEP-FEMU} platform are available at \url{https://github.com/esl-epfl/x-heep-femu} and \url{https://github.com/esl-epfl/x-heep-femu-sdk}.}
    \end{itemize}

    The remainder of this paper is organized as follows. Section~2 reviews related work. Section~3 introduces the \mbox{FEMU} framework. Section~4 describes the \mbox{X-HEEP-FEMU} platform. Section~5 presents three case studies, and, finally, in Section~6 we draw our conclusions.

\section{STATE-OF-THE-ART}

    \begin{table*}[tp]
        \caption{Comparison of relevant FPGA-based platforms across key features.}
        \label{tbl:sota}
        \centering
        \resizebox{0.995\textwidth}{!}{
        \tiny
        \begin{tabular}{|l|c|c|c|c|c|}
        \toprule
        \bf{FPGA Platforms} & \bf{HS-based RH} & \bf{OS-based CS} & \bf{IP Virtualization} & \bf{Performance Estimation} & \bf{Energy Estimation} \\
        \midrule
        LiME~\cite{lime}    & \textcolor{myred}{\xmark} & \textcolor{myred}{\xmark} & \textcolor{myred}{\xmark} & \textcolor{mygreen}{\cmark} & \textcolor{myred}{\xmark} \\
        Hybrid~\cite{hybrid}  & \textcolor{myred}{\xmark} & \textcolor{mygreen}{\cmark} & \textcolor{mygreen}{\cmark} & \textcolor{mygreen}{\cmark} & \textcolor{myred}{\xmark} \\
        FAME~\cite{fame}    & \textcolor{myred}{\xmark} & \textcolor{mygreen}{\cmark} & \textcolor{myred}{\xmark} & \textcolor{mygreen}{\cmark} & \textcolor{myred}{\xmark} \\
        Extrapolator~\cite{extrapolator} & \textcolor{myred}{\xmark} & \textcolor{mygreen}{\cmark} & \textcolor{myred}{\xmark} & \textcolor{mygreen}{\cmark} & \textcolor{myred}{\xmark} \\
        ULPemu~\cite{ulpemu}    & \textcolor{mygreen}{\cmark} & \textcolor{myred}{\xmark} & \textcolor{myred}{\xmark} & \textcolor{mygreen}{\cmark} & \textcolor{mygreen}{\cmark} \\
        ACE~\cite{ace_p}   & \textcolor{myred}{\xmark} & \textcolor{mygreen}{\cmark} & \textcolor{myred}{\xmark} & \textcolor{mygreen}{\cmark} & \textcolor{myred}{\xmark} \\
        SnifferSoC~\cite{sniffersoc}   & \textcolor{myred}{\xmark} & \textcolor{myred}{\xmark} & \textcolor{myred}{\xmark} & \textcolor{mygreen}{\cmark} & \textcolor{mygreen}{\cmark} \\
        ThermalMPSoC~\cite{thermalmpsoc} & \textcolor{myred}{\xmark} & \textcolor{myred}{\xmark} & \textcolor{myred}{\xmark} & \textcolor{mygreen}{\cmark} & \textcolor{mygreen}{\cmark} \\
        HLL~\cite{hll}   & \textcolor{myred}{\xmark} & \textcolor{myred}{\xmark} & \textcolor{myred}{\xmark} & \textcolor{mygreen}{\cmark} & \textcolor{myred}{\xmark} \\
        HERO~\cite{hero}    & \textcolor{mygreen}{\cmark} & \textcolor{mygreen}{\cmark} & \textcolor{mygreen}{\cmark} & \textcolor{mygreen}{\cmark} & \textcolor{myred}{\xmark} \\
        Plug~\cite{plug}    & \textcolor{mygreen}{\cmark} & \textcolor{myred}{\xmark} & \textcolor{mygreen}{\cmark} & \textcolor{mygreen}{\cmark} & \textcolor{myred}{\xmark} \\
        SoftPower~\cite{softpower}    & \textcolor{mygreen}{\cmark} & \textcolor{myred}{\xmark} & \textcolor{myred}{\xmark} & \textcolor{mygreen}{\cmark} & \textcolor{mygreen}{\cmark} \\
        DAQ~\cite{daq}  & \textcolor{mygreen}{\cmark} & \textcolor{myred}{\xmark} & \textcolor{myred}{\xmark} & \textcolor{myred}{\xmark} & \textcolor{myred}{\xmark} \\
        \midrule
        \textbf{FEMU (this work)} & \textcolor{mygreen}{\cmark} & \textcolor{mygreen}{\cmark} & \textcolor{mygreen}{\cmark} & \textcolor{mygreen}{\cmark} & \textcolor{mygreen}{\cmark} \\
        \bottomrule
        \end{tabular}
        }
    \end{table*}

    This section reviews the most relevant FPGA-based platforms that have inspired the development of the proposed \mbox{FEMU} framework. Table~\ref{tbl:sota} compares these platforms across five key dimensions relevant for TinyAI system prototyping and evaluation:~(1)~HS-based RH, referring to platforms that feature an HS implemented in an RH to enable quick prototyping;~(2)~OS-based CS, indicating platforms equipped with a CS capable of running a standard operating system to handle coordination and communication tasks;~(3)~IP virtualization, which supports flexible integration and rapid prototyping by allowing modules to be emulated in software before hardware deployment;~(4)~performance estimation, which is essential for identifying execution bottlenecks and evaluating system responsiveness; and~(5)~energy estimation, which allows designers to assess and optimize power consumption, a key requirement for battery-powered applications at the edge. The following discussion analyzes these features in descending order of frequency across the surveyed platforms and progressively filters the set of platforms based on feature availability.
    
    The feature most frequently supported is performance estimation, which provides essential insight into execution behavior and system responsiveness. It enables the identification of compute and memory bottlenecks and guides hardware/software optimization through profiling and benchmarking. A broad set of platforms includes this capability. LiME~\cite{lime}, Hybrid~\cite{hybrid}, FAME~\cite{fame}, and Extrapolator~\cite{extrapolator} use memory emulation and latency injection to simulate realistic memory behaviors. ULPemu~\cite{ulpemu} supports fine-grained performance tracing through a cycle-accurate monitoring unit. ACE~\cite{ace_p}, SnifferSoC~\cite{sniffersoc}, and ThermalMPSoC~\cite{thermalmpsoc} integrate runtime counters and activity tracking. HLL~\cite{hll} enables profiling in reconfigurable systems. HERO~\cite{hero}, Plug~\cite{plug}, and SoftPower~\cite{softpower} provide task-level and instruction-level metrics for embedded processing. In contrast, DAQ~\cite{daq} lacks any performance analysis capabilities and is thus excluded from the rest of this analysis.
    
    Among the platforms that support performance estimation, the next filter considers the presence of an HS-based RH. This feature is central to enabling hardware/software co-design and the rapid prototyping of accelerators. Platforms that meet this criterion include ULPemu~\cite{ulpemu}, HERO~\cite{hero}, Plug~\cite{plug} and SoftPower~\cite{softpower}. These systems instantiate soft-core processors, such as Ibex or PULP clusters, within the FPGA, allowing realistic interactions with accelerators. Other platforms, such as LiME~\cite{lime}, FAME~\cite{fame}, and Extrapolator~\cite{extrapolator}, focus primarily on memory or host-side emulation and are excluded from further consideration.
    
    We now narrow our analysis to platforms that, in addition to performance estimation and HS-based RH, also support an OS-based CS. This feature is critical for supporting realistic system execution, allowing multitasking, high-level application frameworks, and standardized development environments. Of the remaining platforms, only HERO~\cite{hero} includes an operating system-capable control subsystem. It integrates a Linux-capable ARM Cortex-A host with a tightly coupled PULP cluster, providing OpenMP support and shared virtual memory abstractions. ULPemu~\cite{ulpemu}, Plug~\cite{plug}, and SoftPower~\cite{softpower} operate in bare metal environments and are excluded from the remaining discussion.
    
    Among platforms satisfying the first three features, we now consider energy estimation, which is essential in TinyAI systems due to strict power constraints. This feature enables designers to assess the energy implications of architectural choices and optimize systems accordingly. Although several platforms, such as ULPemu~\cite{ulpemu}, SoftPower~\cite{softpower}, SnifferSoC~\cite{sniffersoc}, and ThermalMPSoC~\cite{thermalmpsoc}, support energy estimation via switching-activity-based or instruction-level models, they were earlier excluded due to a lack of OS-based CS. HERO~\cite{hero}, the only remaining platform, does not provide native support for energy estimation. Its toolchain lacks power models and does not include energy profiling infrastructure. As a result, no platform remains in the filtered set beyond this point.
    
    Lastly, IP virtualization allows developers to emulate hardware components in software, enabling incremental system development and debugging. Although HERO~\cite{hero}, Hybrid~\cite{hybrid}, and Plug~\cite{plug} offer virtualization mechanisms, e.g., through shared memory and software-hardware interface, it was shown that they lack other relevant characteristics such as energy estimation strategies.
    
    In summary, while many platforms offer partial support for the key features required for TinyAI system exploration, none of the surveyed solutions provide full support for all five. This analysis highlights a significant gap in the current landscape and motivates the development of the proposed \mbox{FEMU} platform, which is the only solution in Table~\ref{tbl:sota} to simultaneously support all five dimensions. By integrating reconfigurable hardware heterogeneity, operating system abstraction, energy modeling, performance profiling, and virtualization support, \mbox{FEMU} provides a complete and extensible infrastructure for end-to-end TinyAI HS prototyping.

\section{THE FRAMEWORK}

    \begin{figure} [t]
        \centering
        \includegraphics[width=0.49\textwidth]{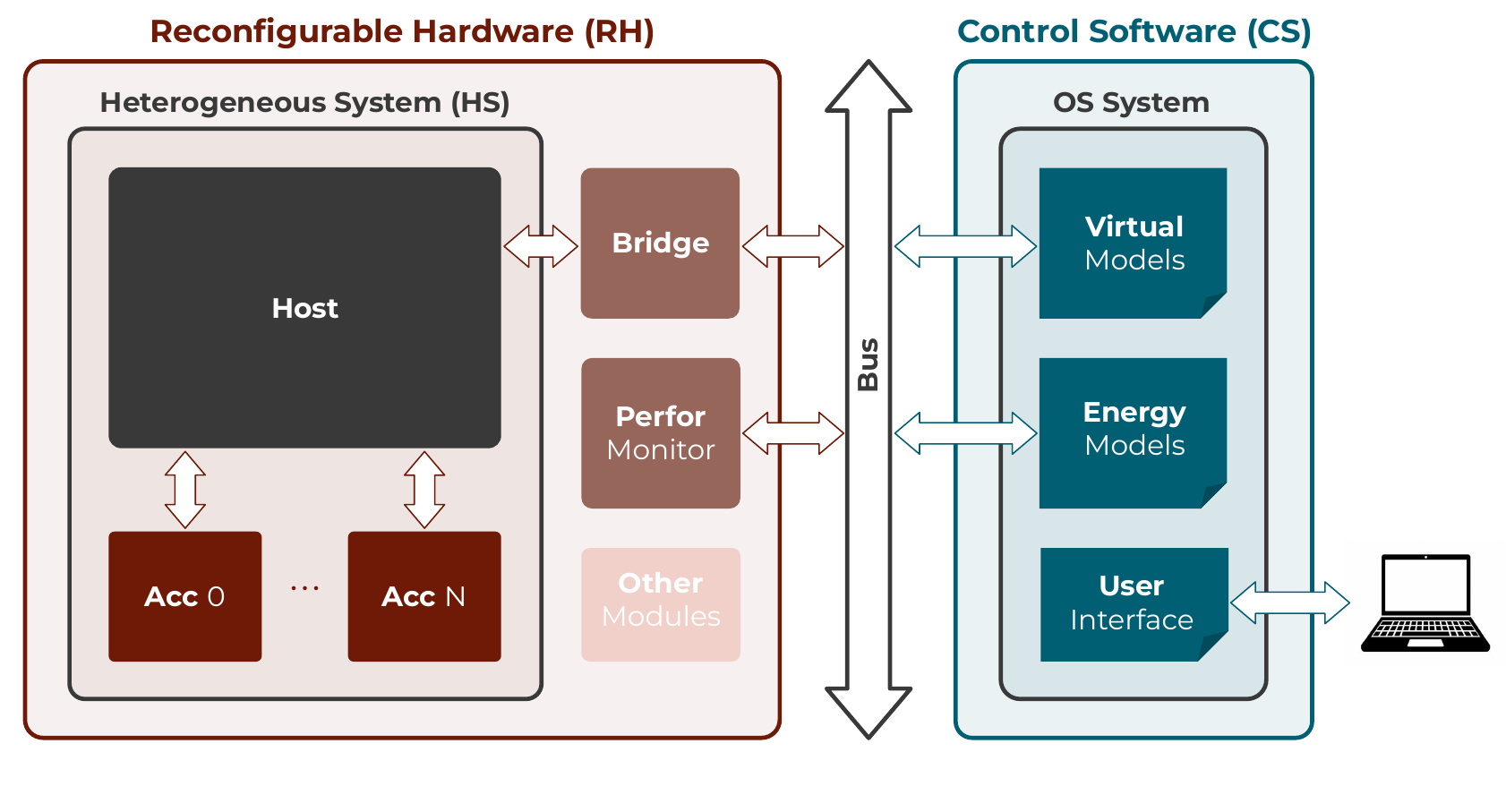}
        \caption{Block diagram of the \mbox{FEMU} framework for prototyping of TinyAI heterogeneous systems.}
        \label{fig:femu_blocks}
    \end{figure}

    This section introduces the structure of the \mbox{FEMU} framework, highlighting its fundamental components and their interactions. Then, it proposes a process to use the framework to prototype and evaluate TinyAI HS.

    \subsection{Architecture and Components}
        
        As illustrated in Figure~\ref{fig:femu_blocks}, the framework comprises two main components: an RH region, which includes hardware modules, and a CS region, which provides the software infrastructure for control, development, and evaluation. The under-development HS is implemented in the RH to support rapid prototyping and flexible architectural exploration of both the host and the connected accelerators. The host features a CPU, memory banks, a bus, and peripherals, which enable internal functionality and external communication. 
        
        In addition to the core HS, the RH integrates additional modules, including a bridge to connect the HS to the CS to support virtualization, and a performance monitor to collect runtime metrics during application execution. Complementing the RH, the CS runs a standard OS on an application-class CPU, enabling the execution of high-level software responsible for key functionalities such as virtualization, energy estimation, flexible user interaction, and test automation.
        
        Virtualization is a key feature of the framework, providing software-driven abstractions for essential system components, including the debugger, ADC, flash, and accelerators. These capabilities decouple software development from hardware implementation, enabling flexible exploration, faster iteration cycles, and reduced development time.
        
        Debugger virtualization allows the CS to fully control the HS under test and the applications running on it, eliminating the need for external programmers or debuggers during development and enabling full test automation. This significantly simplifies the software development and debugging process, especially in early-stage prototyping, by allowing seamless reprogramming, execution control, debugging, and automation of a batch of tests directly from a script.
        
        ADC virtualization allows the system to emulate real-time sensor acquisition by streaming pre-recorded datasets from memory at configurable sampling rates. To implement this functionality, two circular buffers are employed: the first, implemented in software, transfers samples from a large external storage to the CS memory; the second, implemented in hardware, moves data from the CS memory to the RH memory, making it available to the HS at the desired acquisition rate. This mechanism enables an accurate evaluation of performance and energy consumption during the acquisition phase under realistic timing constraints, without requiring physical sensors or external ADCs. In addition, it supports end-to-end validation of the entire system using full datasets, ensuring functional correctness under representative operating conditions.
        
        Flash virtualization provides a flexible and high-performance abstraction of non-volatile storage, enabling the system to read and write large volumes of data directly from the CS memory. This approach removes the latency and bandwidth bottlenecks of physical flash devices, allowing fast and repeated access to large datasets without hardware limitations. By supporting both read and write operations, virtualized flash enables full-cycle experimentation across a wide range of use cases. For example, data-intensive applications benefit from the ability to stream large input datasets quickly; algorithm validation is accelerated by the ease of injecting test vectors and storing intermediate results; and machine learning workflows, such as training and inference, are simplified by supporting fast, iterative access to data and logging outputs.
        
        Finally, accelerator virtualization enables the integration of hardware accelerators as software models within the CS for early-stage prototyping. This approach allows functional validation even before the corresponding RTL implementation is available. By simulating an accelerator, developers can assess the impact of offloading specific tasks to hardware and fine-tune software-accelerator interactions early in the design process. Moreover, it helps decouple the work of software and hardware teams during initial development phases, allowing both to progress in parallel and reducing integration bottlenecks later in the design flow.
        
        Together, these virtualization capabilities enable rapid prototyping, simplify both development and debugging, and foster a seamless transition from high-level software models to hardware implementations.
        
        To enable energy-aware development, the framework includes energy estimation capabilities that combine performance metrics obtained from the RH with technology-specific energy models. These capabilities allow for estimating the energy consumption of applications running on the HS using different combinations of hardware accelerators and software algorithms.
        
        Finally, a user-friendly interface facilitates system control and interaction, enabling users to leverage the capabilities of the framework fully. 
        

        \begin{figure*} [t]
            \centering
            \includegraphics[width=1\textwidth]{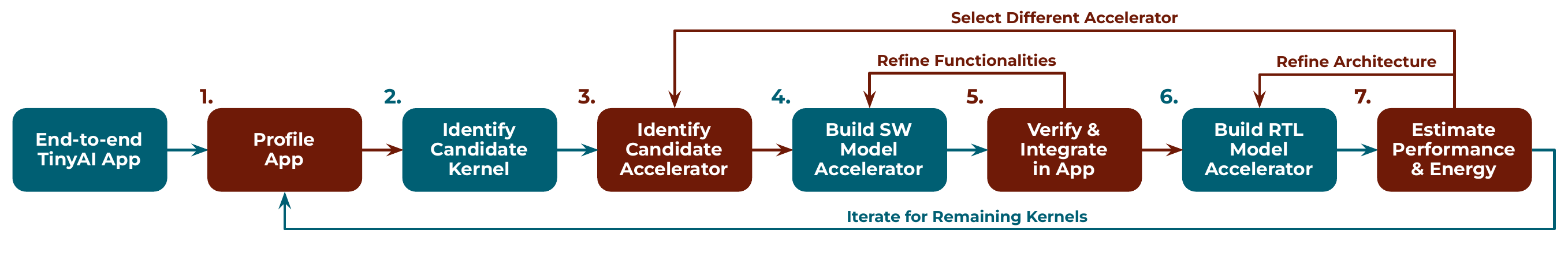}
            \caption{Design cycle for prototyping of TinyAI accelerators.}
            \label{fig:femu_flow}
        \end{figure*}

    \subsection{Prototyping and Evaluation Flow}
    
        Figure~\ref{fig:femu_flow} illustrates the iterative prototyping and evaluation process enabled by the \mbox{FEMU} framework. This flow is designed to support the systematic identification, integration, and refinement of hardware accelerators for TinyAI applications, while tightly coupling functional correctness, performance analysis, and energy estimation.
    
        The process begins with the development or selection of an end-to-end TinyAI application, which is initially executed entirely on the CPU integrated in the host subsystem within the RH and managed via the bridge logic. This CPU-only run is profiled using the performance monitor and energy models provided by the CS to establish a baseline in terms of latency and energy consumption (Step~1).
    
        Based on the profile results, the developer analyzes the application to identify computationally intensive kernels (Step~2). These kernels serve as candidates for offloading to specialized hardware accelerators.
    
        Once a target kernel is selected, the developer identifies a suitable candidate accelerator (Step~3) and builds a corresponding high-level software model (Step~4). This model is executed and validated within the CS, ensuring functional correctness and integration fidelity by comparing its output with the baseline CPU-only implementation (Step~5).
    
        Upon successful validation, the software model is transformed into an RTL implementation using high-level synthesis (HLS) or a hardware description language such as Verilog or VHDL (Step~6). The RTL accelerator is then integrated into the RH along with the host and connected via the shared infrastructure. Its performance is evaluated using the built-in monitoring infrastructure (Step~7). In parallel, the design undergoes synthesis, place-and-route, and power analysis using standard EDA tools to derive a technology-specific energy model. This energy model is combined with the host's energy profile to form a unified estimation framework that allows an accurate evaluation of the energy consumed by the kernel when executed on the accelerator and facilitates comparison against the CPU-only baseline (Step~7).
        
        This structured and iterative flow allows developers to progressively refine accelerator designs to meet specific performance and energy objectives while ensuring functional correctness through end-to-end validation on representative datasets. The cycle can be repeated to refine the accelerator software-level functionality or RTL architecture, explore alternative accelerator candidates, or target additional kernels. By integrating all stages, profiling, modeling, integration, and evaluation, within a unified environment, \mbox{FEMU} enables fast, accurate, and scalable exploration of heterogeneous TinyAI systems.

\section{AN IMPLEMENTATION} \label{sec:impl}

\begin{figure*} [t]
            \centering
            \includegraphics[width=1\textwidth]{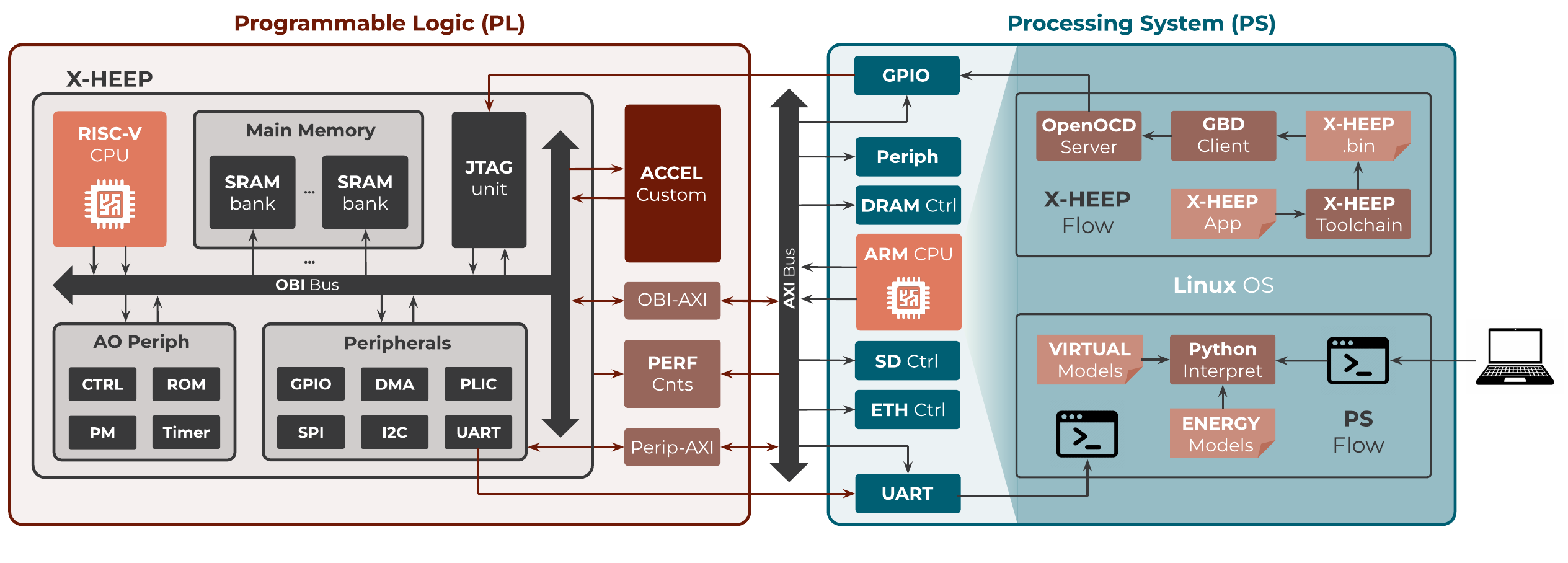}
            \caption{Block diagram of the \mbox{X-HEEP-FEMU} platform for rapid prototyping of TinyAI heterogeneous systems, implemented on the Xilinx Zynq-7020 on the Pynq-Z2 board.}
            \label{fig:x-heep_femu}
        \end{figure*}

    
    This section presents an instantiation of the \mbox{FEMU} framework using real-world hardware and software components, detailing the implementation process. Designers can follow the proposed flow to create custom instantiations of the framework tailored to their selected components. Figure~\ref{fig:x-heep_femu} illustrates the overall block diagram of the case study, referred to as \mbox{X-HEEP-FEMU}.

    \subsection{Choosing an SoC-Based FPGA and Host}

        The initial step of the process involves selecting a suitable SoC-based FPGA to accommodate both the HS and the supporting software infrastructure. To this end, we choose the Xilinx Zynq-7020 SoC integrated on the Pynq-Z2 development board, as it combines the following essential components:~(1)~a programmable logic (PL) region as RH where the HS and additional hardware modules can be flexibly implemented; and~(2)~an ARM Cortex-A9 processing system (PS) region as CS capable of running a Linux-based OS. \mbox{Ubuntu} is chosen as the OS for the PS region, as it provides a lightweight and customizable Linux environment that is specifically suitable for resource-constrained systems. Finally, the Pynq-Z2 board is very affordable, which reduces entry-level costs.

        To support OS execution, the Pynq-Z2 board also includes:~(1)~a secure digital (SD) card reader to load a bootable \mbox{Ubuntu} image;~(2)~\SI{512}{\mega\byte} of dynamic random access memory (DRAM) for the OS and user applications; and~(3)~an Ethernet port for network connectivity, enabling remote access and control, and eliminating the need for physical board access during development.

        After selecting a suitable SoC-based FPGA, choosing a configurable and extensible host is essential for connecting accelerators with a wide range of requirements. To this end, we choose the \mbox{X-HEEP}~\cite{x-heep_1} host due to its modular and flexible architecture, \mbox{RISC-V} compatibility, and suitability for the TinyAI domain.

    \subsection{Setting up a Virtualization Strategy}
        
        After selecting an SoC-based FPGA and a host, we define a virtualization strategy to replicate the functionalities of key system components, including the debugger, ADC, flash, and accelerators. Each element is virtualized through a combination of hardware connections and software abstractions, allowing full control and interaction from the CS without requiring external connectivity.
        
        For debugger virtualization, the \mbox{X-HEEP} JTAG unit is connected to the GPIO ports exposed by the PS. This interface is controlled using a combination of the GNU Debugger (GDB) as the client and OpenOCD as the server. OpenOCD drives the GPIO ports to implement the JTAG protocol, allowing complete control over \mbox{X-HEEP} directly from the \mbox{Ubuntu} environment. In addition, the \mbox{X-HEEP} UART is routed to a UART port exposed by the PS, enabling application-level logging and debugging through standard terminal interfaces.

        To implement ADC virtualization, an SPI-to-AXI bridge is instantiated in the PL to translate the SPI requests generated by \mbox{X-HEEP} into AXI read transactions. This bridge is memory-mapped onto the AXI bus of the PS, allowing straightforward configuration and control. To support real-time acquisition, the bridge is connected to a hardware first-in, first-out (FIFO) buffer that preloads DRAM samples by accessing an AXI port exposed by the PS. In parallel, a second FIFO, implemented in software within the PS, transfers samples from the off-board SD card to the on-board DRAM. This dual FIFO mechanism ensures that data samples are readily available when requested by the HS, meeting timing requirements without incurring additional memory access latency that could otherwise distort performance and energy measurements.

        Flash virtualization is implemented by connecting a second instance of the SPI-AXI bridge to an AXI slave port exposed by the PS, enabling direct access to DRAM. Unlike in ADC virtualization, the bridge supports both read and write operations, enabling not only high-speed streaming of data into the HS but also the storage of processed results or logging data back into memory.

        To implement accelerator virtualization, \mbox{X-HEEP} writes configuration parameters and input data to predefined DRAM regions through an OBI-AXI bridge, accessing the AXI ports exposed by the PS. The accelerator software model running on the PS monitors these memory regions, executes the required computations, and writes the results back to the same memory space. This allows the software model to behave as if it were integrated in the PL alongside the host, ensuring seamless communication and synchronization with the HS under test.
        


    \subsection{Setting up a Performance Estimation Strategy}

        To support the performance evaluation of the applications, dedicated performance counters are integrated into the PL alongside \mbox{X-HEEP}. These counters monitor the activity of various \mbox{X-HEEP} domains by tracking control signals such as clock enable, power enable, and memory state. Specifically, they measure the number of clock cycles that each domain spends in different power states:~(1)~active,~(2)~clock-gated,~(3)~power-gated, and~(4)~retention (for memories).

        Performance counters operate in two modes:~(1)~automatic mode, in which they are automatically activated at the beginning and end of application execution with no user intervention; and~(2)~manual mode, where the application explicitly toggles a dedicated GPIO signal to start and stop the measurement. The manual mode enables developers to profile specific regions of interest within the code, allowing fine-grained performance analysis and targeted optimization. Configuration registers and performance counter values are memory-mapped on the PS bus for straightforward access.

        When accelerators are instantiated as RTL modules in the PL, additional performance counters can be included to monitor their activity and power-state transitions. This allows for the estimation of accelerator-level performance metrics in a manner consistent with the rest of the system. 
        



    \subsection{Setting up an Energy Estimation Strategy}
    
        
        To provide energy estimation, an energy model is derived from a TSMC \SI{65}{\nano\meter} CMOS silicon implementation of \mbox{X-HEEP}, called \mbox{HEEPocrates}~\cite{heepocrates}, and specifies the average power consumption of each domain in its four power states:~(1)~active,~(2)~clock-gated,~(3)~power-gated, and~(4)~retention (for memories). Energy consumption is calculated by multiplying the average power values by the time spent in each state, as measured by the performance counters. The energy consumed by each domain is then summed to obtain the total energy used during application execution. These calculations are performed by the PS after retrieving the performance data at the end of the execution.

        When accelerators are integrated as RTL implementations in the PL, user-defined energy models estimate their individual contributions and incorporate them into the overall system-level evaluation.

        

        \begin{figure*} [t]
            \centering
            \includegraphics[width=1\textwidth]{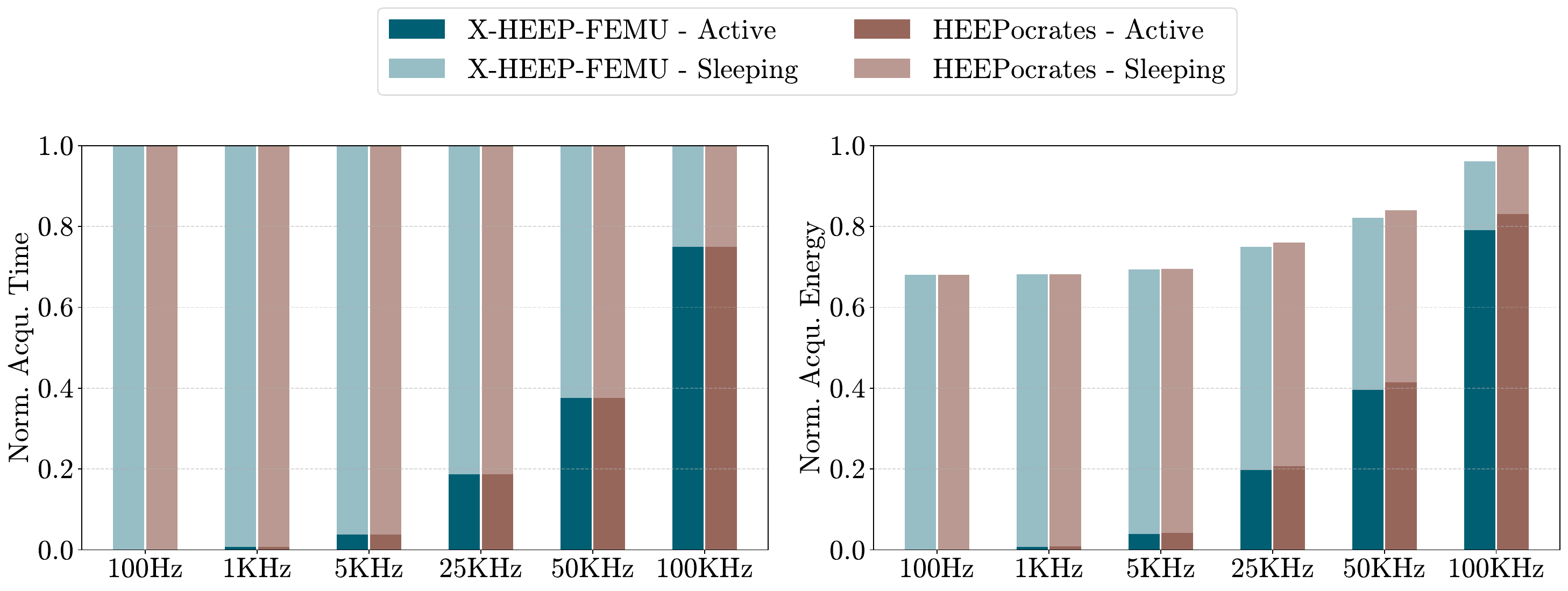}
            \caption{Normalized acquisition time and energy on the \mbox{X-HEEP-FEMU} platform and the \mbox{HEEPocrates} chip for a \SI{5}{\second} acquisition window, with sampling frequency varying from \SI{100}{\hertz} to \SI{100}{\kilo\hertz}.}
            \label{fig:acquisition}
        \end{figure*}

    \subsection{Setting up a User Interface}

        The final step involves defining a flexible user interface for the platform. To this end, a Python class is developed to provide streamlined access to the platform functionalities, eliminating the need to understand low-level hardware and software details. 

        To enhance the user experience, the Python class is integrated with Jupyter Notebooks, allowing users to interact with the platform through a web browser. This enables any HTTP client to connect to the platform and access its internal functionalities intuitively and interactively.
        

\section{CASE STUDIES} 

    To validate the capabilities of the proposed \mbox{FEMU} framework and its instantiation as the \mbox{X-HEEP-FEMU} platform, we define a set of experiments covering the following use cases:~(1)~signal acquisition characterization,~(2)~computation of typical TinyAI workloads,~(3)~sample collection and storage. The objective is to demonstrate that \mbox{X-HEEP-FEMU} enables realistic prototyping and evaluation of performance and energy across various real-world scenarios.

    \subsection{Signal acquisition characterization}

        In this use case, a kernel is configured to run on the \mbox{X-HEEP} CPU and acquire a \SI{5}{\second} window of pre-sampled data using the SPI peripheral. Data is retrieved from DRAM via the ADC virtualization mechanism described in Section~\ref{sec:impl}, which emulates the behavior of a real ADC without requiring physical connectivity.

        To emulate different application scenarios, the kernel is executed in six sampling frequencies ranging from \SI{100}{\hertz} to \SI{100}{\kilo\hertz}. Execution time is measured using the performance counters integrated in the PL, while energy consumption is estimated through the energy model introduced in Section~\ref{sec:impl}.

        As a baseline for validation, the same acquisition kernel is executed on \mbox{HEEPocrates}~\cite{heepocrates}, a silicon implementation of \mbox{X-HEEP} fabricated in TSMC \SI{65}{\nano\meter} CMOS technology and running at \SI{20}{\mega\hertz}, \SI{0.8}{\volt}. In this setup, the pre-sampled data is stored in the on-board flash memory and accessed via the SPI interface. Execution time is measured using CPU performance counters, and energy is estimated by measuring the average power during acquisition using on-board instrumentation.

        Figure~\ref{fig:acquisition} presents the normalized acquisition time and energy for a \SI{5}{\second} acquisition window across the selected sampling frequencies. The results are reported for both the \mbox{X-HEEP-FEMU} platform and the \mbox{HEEPocrates} chip, highlighting the contribution of active and sleeping periods for each configuration.

        At low sampling frequencies, both platforms spend the vast majority of the time in sleep mode, with active periods contributing less than \SI{1}{\percent} to the total acquisition time and energy. As the sampling frequency increases, the proportion of active time increases significantly, and the system transitions to an active-dominated regime, where the active phase accounts for more than \SI{70}{\percent} of the total acquisition time and energy. This trend offers valuable insight for system designers, enabling them to focus architectural optimizations on either the active or sleep phase, depending on the target acquisition frequency.

        \begin{figure*} [t]
            \centering
            \includegraphics[width=1\textwidth]{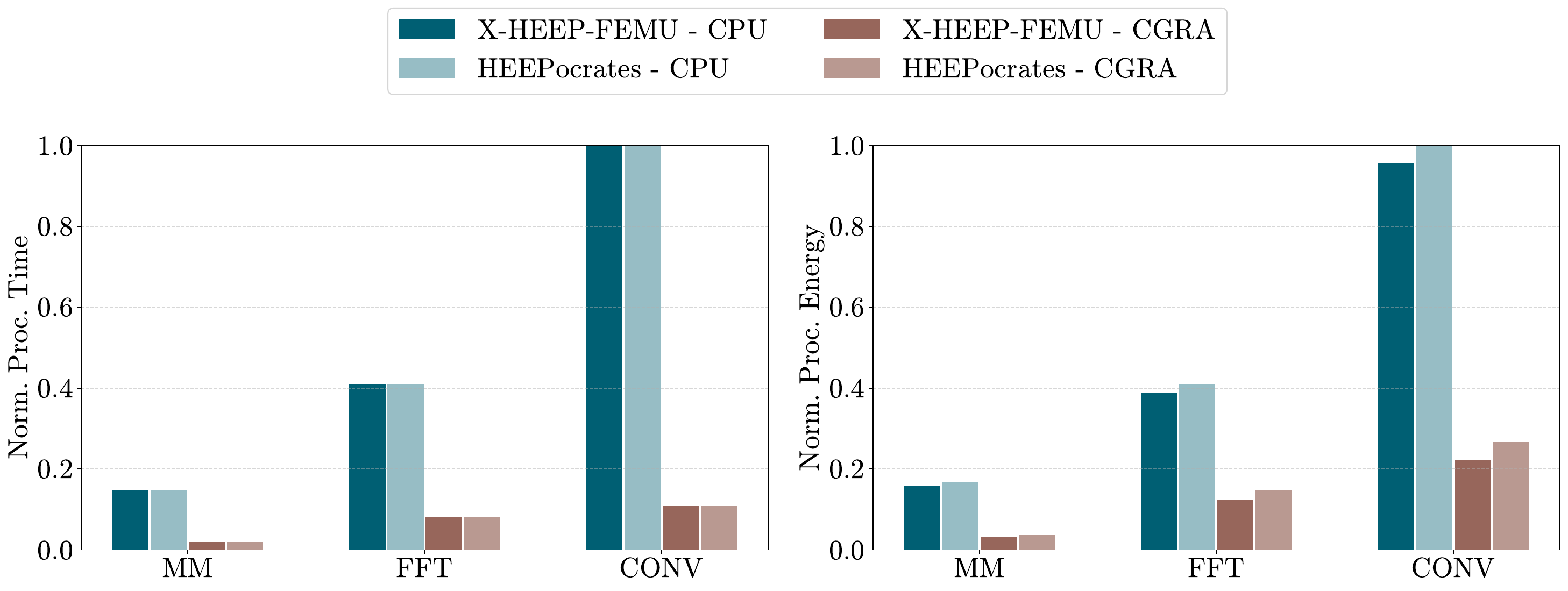}
            \caption{Normalized processing time and energy on the \mbox{X-HEEP-FEMU} platform and the \mbox{HEEPocrates} chip for three representative kernels executed on both the CPU and the CGRA accelerator.}
            \label{fig:processing}
        \end{figure*}

    \subsection{Computation of typical TinyAI workloads}

        In this use case, three representative kernels from typical TinyAI workloads are evaluated: a 121x16 by 16x4 matrix multiplication (MM) in INT32~\cite{najafi2024versasens}, a 2D convolution (CONV) with a 16x16 input with three channels and eight filters of 3x3, also in INT32~\cite{carpentieri2024performance}, and a 512-point fast Fourier transform (FFT) in FxP32~\cite{vwr2a}. Each kernel is executed in two configurations:~(1)~on the \mbox{X-HEEP}~\cite{x-heep_2} CPU as a baseline and~(2)~accelerated using a coarse-grained reconfigurable array (CGRA)~\cite{oe-cgra}. The CGRA is first modeled in Python~\cite{aspros2025flexible} and virtualized in the PS for early-stage behavioral verification and software development, and then instantiated in the PL for performance and energy estimation. 
        

        In both the CPU-only and RTL-based CGRA versions, execution time is measured using performance counters integrated in the PL, and energy consumption is estimated using the energy model introduced in Section~\ref{sec:impl}. The activity tracking and energy model are extended to support the CGRA, incorporating performance counters and power values derived from post-place and route analysis.

        As a reference baseline, the same processing kernels are also executed on the CPU and CGRA of \mbox{HEEPocrates}, allowing a direct comparison of performance and energy efficiency.

        Figure~\ref{fig:processing} presents the normalized processing time and energy for the selected TinyAI kernels. The results are shown for both the \mbox{X-HEEP-FEMU} platform and the \mbox{HEEPocrates} chip, each evaluated in CPU-only and CGRA-accelerated configurations.

        CGRA acceleration leads to substantial reductions in processing time (up to \SI{9}{\times}) in all workloads. Among the benchmarks, CONV exhibits the largest performance gain due to its high computational intensity and regular memory access pattern, both of which are well-suited to spatial acceleration.

        CGRA execution consistently reduces energy consumption, confirming that acceleration enhances not only performance but also energy efficiency. The energy trends observed across the two platforms are closely aligned: \mbox{X-HEEP-FEMU} accurately replicates the behavior of \mbox{HEEPocrates}, with an average deviation of approximately \SI{5}{\percent} in CPU-only configurations, consistent with the simplified energy model discussed previously. In CGRA-accelerated configurations, the deviation increases to around \SI{20}{\percent}, primarily because the power estimations for CGRA are derived from post-place and route results, which are inherently less accurate than silicon measurements. Despite these discrepancies, consistent energy trends across platforms validate the use of \mbox{X-HEEP-FEMU} for early-stage evaluation of hardware accelerators.

    \subsection{Sample collection and storage}

        Another illustrative use case of the proposed virtualization approaches is presented in~\cite{moisture}, which explores the problem of wood moisture classification. In this application, the flash virtualization mechanism of the \mbox{X-HEEP-FEMU} platform is leveraged to acquire 35000 samples of \SI{16}{\bit} ultra-sound data per window, amounting to over \SI{70}{\kibi\byte} per acquisition. Thanks to this virtualization, each window is transferred in approximately \SI{10}{\milli\second}, compared to the \SI{2.5}{\second} required when using a physical SPI flash memory. As a result, transferring 240 acquisition windows of a complete experiment takes only \SI{2.4}{\second}, down from \SI{10}{\minute}, yielding a \SI{250}{\times} speedup. This demonstrates the effectiveness of flash virtualization in enabling rapid, large-scale data acquisition for real-time machine learning applications.

\section{Conclusion}

    In this paper, we have introduced \mbox{FEMU}, a new open-source and configurable emulation framework for rapid prototyping and evaluation of TinyAI HS. \mbox{FEMU} exploits the capabilities of SoC-based FPGAs to seamlessly integrate an RH region, where under-development HS designs can be prototyped, with a CS region, running a standard OS environment for system supervision and communication.


    We validated our approach by implementing the \mbox{X-HEEP-FEMU} platform, an instantiation of \mbox{FEMU} on the Xilinx Zynq-7020 SoC. This platform integrates the \mbox{X-HEEP} host in the RH, a Linux-based Python environment on the ARM Cortex-A9 CS, and energy models based on a TSMC \SI{65}{\nano\meter} silicon implementation of \mbox{X-HEEP}, referred to as \mbox{HEEPocrates}. The experimental results confirm that \mbox{X-HEEP-FEMU} provides a practical, accurate, and flexible infrastructure for early-stage design exploration of TinyAI heterogeneous systems, with an energy model exhibiting an error within \SI{5}{\percent}.

\section{Acknowledgments}

    The authors thank Saverio Nasturzio for his valuable contribution in implementing the debugger and flash virtualization infrastructure.
    
    This work was supported in part by the Swiss State Secretariat for Education, Research, and Innovation (SERI) through the SwissChips research project; by the Swiss NSF, grant no. 10.002.812: ``Edge-Companions: Hardware/Software Co-Optimization Toward Energy-Minimal Health Monitoring at the Edge''; and by a donation of material from AMD Xilinx to EPFL.

\printbibliography

\end{document}